\documentclass[prb,twocolumn,superscriptaddress]{revtex4}
\usepackage{graphicx}
\usepackage{epsfig}
\usepackage{amssymb,amsmath,amsfonts,hyperref}
\usepackage{wasysym}
\usepackage{latexsym}
\usepackage{eucal}
\usepackage{verbatim}
\usepackage[normalem]{ulem}
\usepackage{color}

\newcommand{\be}{\begin{equation}}
\newcommand{\ee}{\end{equation}}

\newcommand{\bea}{\begin{eqnarray}}
\newcommand{\eea}{\end{eqnarray}}

\begin{document}

\title{Resonating valence-bond physics on the honeycomb lattice}
\author{Pranay Patil}
\affiliation{{\small Dept. of Physics, Indian Institute of Technology Madras, Chennai 600036}}
\author{Ishita Dasgupta}
\affiliation{{\small Dept. of Physics, Indian Institute of Technology Bombay, Mumbai 400076}}
\author{Kedar Damle}
\affiliation{{\small Dept. of Theoretical Physics, Tata Institute of Fundamental Research, Mumbai 400005}}
\begin{abstract}
We study bond and spin correlations  of the nearest-neighbour resonating valence bond (RVB) wavefunction for a
SU($2$) symmetric $S=1/2$ antiferromagnet on the honeycomb lattice. We find that spin correlations
in this wavefunction are short-ranged, while the
bond energy correlation function takes on an oscillatory power-law form $D(\vec{r}) \sim \cos({\mathbf Q}\cdot {\vec{r}}) /|{\vec{r}}|^{\eta_w(2)}$, where ${\mathbf Q} = (2\pi/3, -2\pi/3)$ is the wavevector corresponding to ``columnar'' valence-bond
solid order on the honeycomb lattice, and $\eta_w(2) \approx 1.49(3)$.
We use a recently introduced large-$g$
expansion approach to relate bond-energy correlators of the SU($g$) wavefunction
to dimer correlations of an interacting fully-packed dimer model
with a three-dimer interaction of strength $V(g)=-\log(1+1/g^2)$. Putting $g=2$, we find
numerically that the dimer correlation function $D^{d}(\vec{r})$ of this
dimer model has power-law behaviour $D^{d}(\vec{r}) \sim \cos({\mathbf Q}\cdot {\vec{r}}) /|{\vec{r}}|^{\eta_d(2)}$ with $\eta_d(2) \approx 1.520(15)$, in rather good agreement with the wavefunction results. We also study the same quantities for $g=3,4,10$ and find that the
bond-energy correlations in the SU($g$) wavefunction are consistently well-reproduced
by the corresponding dimer correlations in the interacting dimer model.
\end{abstract}
\maketitle

\section{Introduction}
As is well-known, localized electronic moments (spins) in Mott-insulating
materials typically interact with near-neighbours via antiferromagnetic exchange
interactions which can be much bigger than the weak magnetic dipole
interactions between these localized moments. The possibility that such quantum antiferromagnets remain in a liquid-like phase down to the lowest temperature has attracted sustained
interest since the early work of Fazekas and Anderson\cite{Fazekas_Anderson}.

This has motivated the study of candidate wavefunctions
that describe various quantum spin liquid ground states. Here, our focus is on
a particular construction that works directly in the overcomplete basis of singlet (valence) bonds
between spins, by specifying amplitudes for various ways in which the spins
can pair up to make singlets. The full ``resonating valence bond'' (RVB) wavefunction is then a superposition of all these possibilities, with these amplitudes
chosen by some physically motivated rule.
On bipartite lattices, it is possible to choose the phase of these amplitudes so as to satisfy the
Marshall sign-rule\cite{Marshall}, which is known to be obeyed in the ground
state of a large class of antiferromagnets. Indeed, in their
original study of such RVB wavefunctions on the square lattice, Liang, Doucot
and Anderson\cite{Liang_Doucot_Anderson} had fixed the sign-structure in this manner
to study the variational energy of the square-lattice $S=1/2$ Heisenberg antiferromagnet as a function of the length distribution of valence bonds. They concluded that the short-ranged
RVB wavefunction with valence-bonds allowed only between pairs
of nearest-neighbour spins gives a variational energy
that is only slightly higher
than that of a trial N\'eel ordered state, which has long-ranged valence-bonds in this basis.

More recent work has built on these results in several ways: First,
Tang, Sandvik and Henley\cite{Tang_Sandvik_Henley} and Albuquerque and Alet\cite{Albuquerque_Alet}
revisited the square lattice nearest-neighbour RVB (nnRVB) wavefunction 
using the loop algorithm of Sandvik and Evertz\cite{Sandvik_Evertz} to sample expectation
values in the nnRVB wavefunction. They found that bond-energy correlations have a slow, oscillatory
power-law decay. This is in complete contrast to the extremely short-ranged spin correlations in this wavefunction. Second, Cano and Fendley\cite{Cano_Fendley} constructed a spin Hamiltonian with short-ranged
couplings whose ground state is the nnRVB wavefunction on the square lattice.
Third, one of the present authors, in collaboration with Dhar and Ramola, developed a cluster-expansion approach that
relates the bond-energy correlations in this square lattice wavefunction to
dimer correlations of a specific interacting dimer model, whose leading interaction
is an attraction between pairs of parallel dimers on adjacent bonds of the square lattice\cite{Damle_Dhar_Ramola}. Some aspects of this correspondence were also checked 
by St\'ephan {\em et. al.}\cite{Stephan_Ju_Fendley_Melko} in their study of the entanglement properties
of this wavefunction. 
In the three-dimensional case, Albuquerque, Alet and Moessner\cite{Albuquerque_Alet_Moessner} showed that the nnRVB wavefunction on the cubic and diamond lattices has long-range
antiferromagnetic order, underlining the importance of dimensionality
in determining the nature of correlations in this kind of variational wavefunction. Additionally,
recent work by Xu and Beach\cite{Xu_Beach_unpublished} suggests that an anisotropic version of the three-dimensional nnRVB wavefunction
describes interesting spin-liquid behaviour.

In the present study, we focus on extending these results to
antiferromagnets on the honeycomb lattice. Our primary motivation
is to test the correspondence between bond-energy
correlations in the nnRVB wavefunction and dimer correlations of a
fully-packed dimer model with certain interactions, the form of which
we derive here using the methods of Ref.~\onlinecite{Damle_Dhar_Ramola}.
To this end, we study both sides of this correspondence using
Monte-Carlo simulations. Our computational results
rely heavily on a new update-scheme\cite{Damle_unpublished} that allows a more efficient and ergodic
Monte-Carlo sampling of valence-bond configurations when used in conjunction with the Sandvik-Evertz
algorithm\cite{Sandvik_Evertz}. To place both our wavefunction
and dimer model results in the context of a long-wavelength
height-model description\cite{Youngblood_Axe_McCoy,Youngblood_Axe,Blote_Hillhorst,Nienhuis_Hillhorst_Blote,Kondev_Henley_PRB, Zeng_Henley,Raghavan_Henley_Arouh,Fradkin_Huse_Moessner_Oganesyan_Sondhi,Kenyon,Boutillier_Tiliere,Alet_etal_PRE}, we first provide a self-contained account of the correspondence between the height-model stiffness and power-law exponents for dimer and bond-energy correlators, and derive the form of the dependence of the reduced dimer partition function on winding numbers. Although 
both these results are available in earlier literature\cite{Fradkin_Huse_Moessner_Oganesyan_Sondhi,Kenyon,Boutillier_Tiliere,Alet_etal_PRE}, our
formulation of the latter may be of some independent interest since it makes explicit the three-fold symmetry of the honeycomb lattice.

The rest of this article is organized as follows: In Section~\ref{loops}
we establish the correspondence between the nnRVB wavefunction on
the honeycomb lattice and an interacting fully-packed dimer
model on the same lattice, with a specific form of the interactions, which we derive.
In Section~\ref{heights}, we discuss the coarse-grained height-model
description of fully-packed dimers on the honeycomb lattice, paying
particular attention to features that are specific to the honeycomb
lattice case. We derive the dependence of the restricted partition
function on winding number sectors, and summarize the correspondence between
height-model stiffness and various power-law exponents. We also
indicate how this can be carried over to our wavefunction studies.
In Section~\ref{algorithm}, we summarize a new update-scheme\cite{Damle_unpublished} for Monte-Carlo
sampling of valence-bond configurations, and explain how it improves
the ergodicity of our simulations when used in conjunction with the well-known
Sandvik-Evertz algorithm\cite{Sandvik_Evertz}.
In Section~\ref{numerics}, we describe our numerical results
on the nnRVB wavefunction for SU($g$) antiferromagnets (with $g=2,3,4,10$)  and the corresponding interacting dimer
model, and demonstrate that the dimer correlations in the latter provides a rather good account of the bond-energy correlations in the former. We close with a brief discussion of our results in Section~\ref{discussion}.

\section{Valence-bonds and fully-packed interacting dimers}
\label{loops}

\begin{figure}
\includegraphics[width=\hsize]{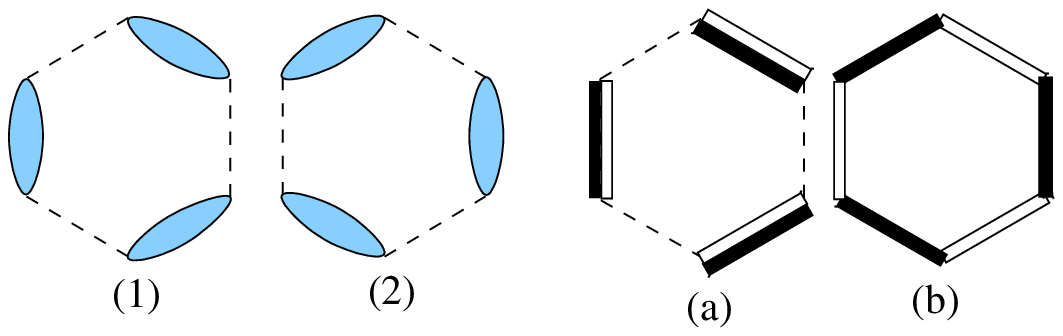}
\caption{$(1)$ and $(2)$ represent the two flippable configurations of dimers around a hexagon (here dimers are represented by ellipses). 
The weight of configuration $(1)$ in the interacting dimer model gets contributions from the two loop configurations
$(a)$ and $(b)$ of the loop model equivalent to the nnRVB wavefunction. This is captured
by the effective interactions between dimers worked out in the text.}
\label{Fig1}
\end{figure}

\begin{figure}
\includegraphics[width=\hsize]{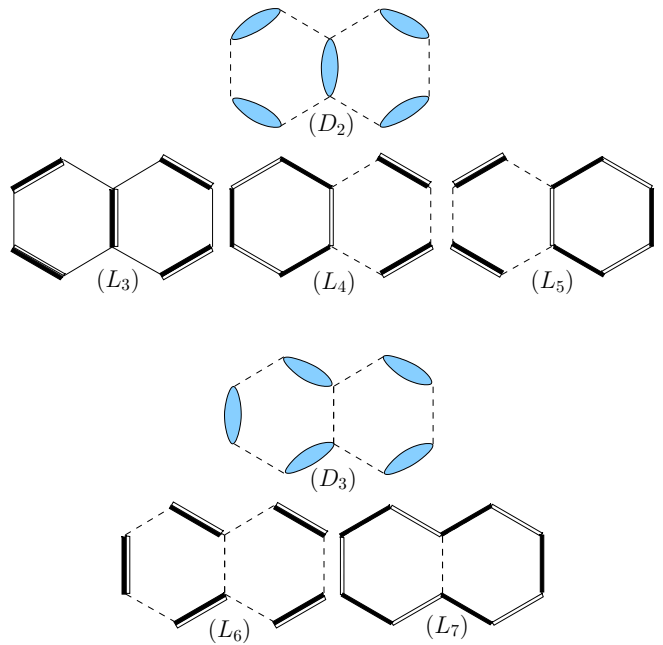}
\caption{$D_2$ shows a configuration with two flippable hexagons that share a dimer. The weight for
such a configuration in the interacting dimer model gets contributions from loop configurations
$L_3$, $L_4$, and $L_5$. Similarly, $D_3$ represents one of the two flippable arrangements of dimers
around on the perimeter of a double-hexagon. The weight for
such a configuration in the interacting dimer model gets contributions from loop configurations
$L_6$, $L_7$, and $L_4$.}
\label{Fig2}
\end{figure}

The nearest neighbour RVB wavefunction for a bipartite SU($2$) antiferromagnet with
$S=1/2$ spins at each site is given by a uniform amplitude superposition
of all possible SU($2$) valence bond solid states in which each spin makes a singlet
with one of its neighbours. This construction generalizes readily to
the SU($g$) case when the SU($g$) spins on one sublattice carry the fundamental
representation and those on the other sublattice carry the complex conjugate
of the fundamental.
In general, we write
\begin{equation}
|\psi(g)\rangle = \sum_{D} |D\rangle_{g}
\end {equation}
where
\begin{equation}
|D\rangle_g = \prod_{l \in D} {|\phi_o(g)\rangle}_l
\end {equation}
where $D$ is any complete dimer cover of the bipartite lattice, 
$l$ denotes all links covered by dimers of $D$, 
$|\phi_o(g)\rangle_l$ is the SU($g$) singlet state of the two spins connected by $l$,
and $A$ and $B$ denote the A and B sublattice sites connected by link $l$.
The norm is given by:
\begin{equation}
\langle\psi(g)|\psi(g)\rangle = \sum_{D,D'} \langle D|D'\rangle_{g}
\end{equation}
which can be written as the partition function of a fully packed loop model with 
non-intersecting loops\cite{Sutherland,Beach_Sandvik,Beach_Alet_Mambrini_Capponi,Lou_Sandvik_Kawashima}.
To see this, one notes that a superposition of two fully-packed dimer-covers
gives a fully-packed configuration of loops. Thus,  
we identify the inner product of $|D\rangle$ and $|D'\rangle$ 
with the loop configuration $L$ whose weight $w_{loop}(g,L)$ is determined
by the quantum-mechanical overlap between these two singlet states.
The norm of the wavefunction can thus be written as the loop gas partition
function
\begin{equation}
Z_{loop}(g)=\sum_{L}w_{loop}(g,L).
\end{equation}
Here, $w_{loop}(g,L)=(g)^{n_d(L)}(2g)^{n_l(L)}$, where $n_d(L)$ is the number
of trivial length-two loops (doubled-edges) in the loop configuration $L$ (corresponding to bra and ket valence bonds occupying the same link) and $n_l(L)$ is the number of non-trivial (length-four or more) loops in $L$.
Operator expectation values in this wavefunction, such as spin correlation and bond-energy
correlation functions, are obtained from Monte-Carlo estimators 
defined entirely in terms of this loop gas\cite{Sutherland,Beach_Sandvik,Beach_Alet_Mambrini_Capponi,Lou_Sandvik_Kawashima}. For instance,
in the SU($2$) case, the two-point correlation function of spins is simply three-fourth of the probability
that both spins lie on the same overlap loop. The estimator for the two-point function of the bond-energy
operator $\vec{S}_i \cdot \vec{S}_j$ corresponding to a nearest-neighbour  bond $\langle ij \rangle$, {\em i.e.}
the correlation of bond-energies at $\langle i j\rangle$ and $\langle k l\rangle$, 
is only slightly more complicated: It involves the probabilities for various ways in which the four
points $i$, $j$, $k$, $l$ lie on at most two loops of the loop gas\cite{Sutherland,Beach_Sandvik,Beach_Alet_Mambrini_Capponi,Lou_Sandvik_Kawashima}. Both these results have natural generalizations\cite{Sutherland,Beach_Sandvik,Beach_Alet_Mambrini_Capponi,Lou_Sandvik_Kawashima} to the SU($g$) case.

As was noted in earlier work\cite{Damle_Dhar_Ramola}, if the loop gas is in a short-loop phase,
corresponding to exponentially decaying spin correlations in the RVB wavefunction, there is a precise mapping between
the loop gas and an interacting fully-packed dimer
model defined on the same bipartite lattice. This mapping is perturbative in $g^{-1}$, 
and leads to a dimer model with $n$-dimer interactions whose strength decays rapidly
with $n$ when $g$ is large. Under this mapping, bond-energy correlators in
the RVB wavefunction map to dimer correlators in this interacting dimer model, apart
from an overall prefactor\cite{Damle_Dhar_Ramola}.
In the square-lattice case, this mapping has proved to be a useful way to understand the power-law
bond-energy correlations in the SU($2$) wavefunction\cite{Damle_Dhar_Ramola}.

Here, we use this approach\cite{Damle_Dhar_Ramola} to derive the form of this interacting dimer model for the honeycomb lattice case. Following Ref.~\onlinecite{Damle_Dhar_Ramola}, the $n$-dimer interaction is given recursively by the following relation :
\begin{equation}
-\log[w_{dimer}(g,D_n)]=V_n(D_n)+\sum^{n-1}_{m=1}\sum_{{D_m}\epsilon{D_n}}{V_m}(D_m),
\end{equation}
where
\begin{equation}
w_{dimer}(g,D_n)=\sum_{L|D_n}\frac{w_{loop}(g,L)}{2^{n_l(L)}},
\end{equation}
where $L|D_n$ denotes all possible loop configurations that contribute to the weight of a given $n$-dimer configuration $D_n$, and $D_m \epsilon D_n$ denotes all sub-configurations
of this $n$-dimer configuration.
Using the above relation, we see that the one-body potential (fugacity) goes as $-\log(g)$. This
simply says that each dimer
contributes a factor of $g$ to the dimer model partition function; since we are considering
a fully-packed dimer model, this just fixes the overall normalization of the partition function.
From this recursion relation, we also see that every two-body term is zero. The only three-body term lives on a flippable hexagon (dimer configurations labeled $(1)$ and $(2)$ in Fig~\ref{Fig1}), which has two possible loop configurations that contribute to the weight of each such flippable dimer configuration; for instance,
loop configurations labeled $a$ and $b$ contribute to the dimer configuration labeled $(1)$ in Fig~\ref{Fig1} (and similarly for $(2)$). Thus, flippable hexagons are
favoured in the interacting dimer model by an attractive three-body term:
\begin{equation}
V_3(g)=-\log(1+g^{-2}).
\end{equation}
Using these results, it is easy to see that the four-body potential is zero.
There are two five-dimer configurations which have a non-zero interaction energy (Fig~\ref{Fig2}). The first,
labeled $D_2$ in Fig.~\ref{Fig2}, gets contributions from loop configurations $L_3$, $L_4$, and $L_5$
as shown in Fig.~\ref{Fig2}. The second, labeled $D_3$, gets contributions from loop configurations $L_4$,  $L_6$ and $L_7$.  Thus, we have the two five-body interaction potentials
\begin{equation}
V_5^a(g)=-\log(1-(g^2+1)^{-2}),
\end{equation}
\begin{equation}
V_5^b(g)=-\log(1+(g^4+g^2)^{-1}).
\end{equation}
From this recursive analysis, it is easy to see that $n$-body terms, when non-zero, are generally of
order $O(g^{-(n-1)})$ for $n >2$. In our computations we only use the leading order non-trivial interaction (which is the three-body term). As we will see in later sections, this already gives
a rather good account of the bond-energy correlators in the SU($g$) wavefunction for $g=2,3,4,10$.

\section{Height model}
\label{heights}

In this section, we lay the groundwork to place both our wavefunction
and dimer model results in the context of a long-wavelength
height-model description\cite{Youngblood_Axe_McCoy,Youngblood_Axe,Blote_Hillhorst,Nienhuis_Hillhorst_Blote,Kondev_Henley_PRB, Zeng_Henley,Raghavan_Henley_Arouh,Fradkin_Huse_Moessner_Oganesyan_Sondhi,Kenyon,Boutillier_Tiliere,Alet_etal_PRE}. 
To this end, we provide a self-contained account of the correspondence between the height-model stiffness and power-law exponents for dimer and bond-energy correlators, and derive the form of the dependence of the reduced partition function on winding numbers. Although 
both these results are available in earlier literature\cite{Fradkin_Huse_Moessner_Oganesyan_Sondhi,Kenyon,Boutillier_Tiliere,Alet_etal_PRE}, our
formulation of the latter may be of some independent interest since it makes explicit the three-fold symmetry of the honeycomb lattice.

\begin{figure}
\includegraphics[width=\hsize]{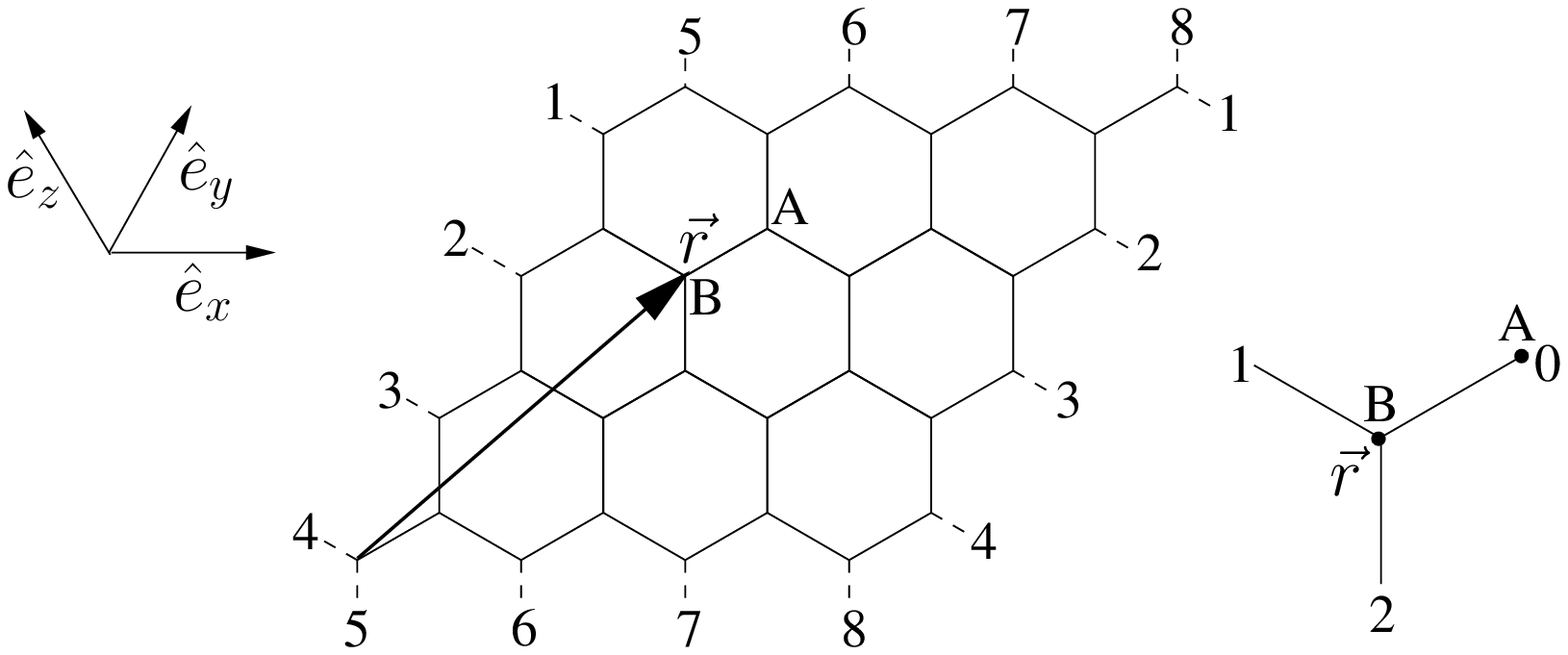}
\caption{The honeycomb lattice is constructed using a two point basis
of sites $A$ and $B$ ``belonging'' to each Bravais lattice point $\vec{r} = m \hat{e}_x
+ n \hat{e}_y$ with integer $m$ and $n$, which, in our convention, is the coordinate of the
$B$-sublattice site. When talking
of bond-energies or dimer occupation numbers, we use the convention that
three types of bonds $0$, $1$, and $2$ ``belong'' to each $\vec{r}$, as shown in the figure. }
\label{Fig3}
\end{figure}

\begin{figure}
\includegraphics[width=\hsize]{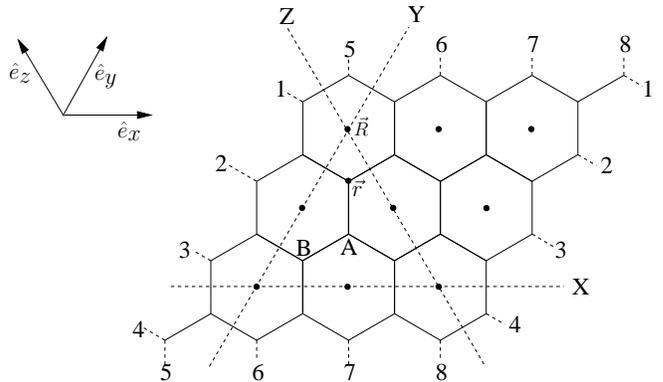}
\caption{$B$-sublattice honeycomb sites are labeled by $\vec{r}$ 
and their corresponding triangular lattice counterparts are labeled
 $\vec{R}$. Also shown here are the lines across which dimer or valence-bond flux is calculated  to get three winding numbers $w_x$, $w_y$ and $w_z$ as described
in the text.}
\label{Fig4}
\end{figure}

We begin by defining a microscopic height field $H(\vec{R})$ on the triangular lattice dual to
the honeycomb lattice (Fig.~\ref{Fig3} and Fig.~\ref{Fig4}) in the following manner: Given a configuration of dimers on the honeycomb lattice, and fixing the height
at the origin of the triangular lattice to be $H(\vec{O})=0$, we construct $H(\vec{R})$ on
sites $\vec{R}$ of the dual triangular lattice using the rules give below
\begin{equation}
H(X+1,Y)-H(X,Y)=n_2(x,y+1)-\frac{1}{3} \; ,
\end{equation}
\begin{equation}
H(X,Y+1)-H(X,Y)=-n_1(x,y+1)+\frac{1}{3} \; ,
\end{equation}
\begin{equation}
H(X-1,Y+1)-H(X,Y)=n_0(x-1,y+1)-\frac{1}{3} \; ,
\end{equation}
where $n_\mu(\vec{r})$ is $1$ if the $\mu^{\rm th}$ type bond belonging to point $\vec{r}$ (Fig.~\ref{Fig3}) is occupied by a dimer and $0$ otherwise. Here, $\vec{r} = (x,y)$ is the coordinate
of a $B$-sublattice site of the honeycomb lattice, we
assign the same coordinate to the $A$-sublattice site ``belonging'' to $\vec{r}$ (as shown in Fig.~\ref{Fig3}),
and 
$\vec{R}=(X,Y)$ is the coordinate of the corresponding dual triangular lattice
site that coincides with the center of the hexagon vertically above this $B$-sublattice
site. Clearly, this microscopic height $H$ is uniquely defined
for all fully-packed configurations, and takes on one-third-integer values.

As is well-understood\cite{Youngblood_Axe_McCoy,Youngblood_Axe,Blote_Hillhorst,Nienhuis_Hillhorst_Blote,Kondev_Henley_PRB, Zeng_Henley,Raghavan_Henley_Arouh,Fradkin_Huse_Moessner_Oganesyan_Sondhi,Kenyon,Boutillier_Tiliere,Alet_etal_PRE}, a fully-packed dimer configuration, and ``nearby'' configurations
accessible to it via local rearrangements, all correspond to the same coarse-grained height-field. Dimer states with many
such nearby configurations give rise to a coarse-grained height-field with small tilt.
If the dimer model is in a power-law ordered state, one expects
that long-wavelength fluctuations of the height-field will be determined by this entropic
cost of tilting the height configuration, while the restriction that $H$ takes on values that are one-third of an integer is expected to be
irrelevant as far as long-distance properties are concerned. One therefore postulates that long-wavelength
properties of such fully-packed dimer models are captured by a 
coarse-grained  {\em real-valued} height field $h$ whose statistics is governed by a Gaussian action\cite{Youngblood_Axe_McCoy,Youngblood_Axe,Blote_Hillhorst,Nienhuis_Hillhorst_Blote,Kondev_Henley_PRB, Zeng_Henley,Raghavan_Henley_Arouh,Fradkin_Huse_Moessner_Oganesyan_Sondhi,Kenyon,Boutillier_Tiliere,Alet_etal_PRE} $S_{\rm renorm.} \sim (\nabla h)^2$, {\em i.e} a statistical weight proportional to
$\exp(-{\rm const.} \int d^2 r (\nabla h)^2)$. 

Here, we prefer a hybrid approach
whereby we work with a {\em real-valued} height field $h(\vec{R})$ that is defined
on the dual triangular lattice sites $\vec{R}$. This is equivalent to re-discretizing the coarse-grained
height action on the dual triangular lattice, instead of imposing an isotropic cutoff in momentum-space. 
Thus, we postulate that a height configuration $h(\vec{R})$
has statistical weight $\exp(-S)$, where $S$ has the form
\begin{equation}
S={\pi \kappa_d}\sum_{\langle \vec{R} \vec{R}' \rangle} (h_{\vec{R}}-h_{\vec{R}'})^2
\end{equation}
where the sum is over all nearest neighbour bonds ${\langle \vec{R} \vec{R}' \rangle}$
of the triangular lattice, and $\kappa_d$ is a phenomenological stiffness constant.

As is well-known\cite{Fradkin_Huse_Moessner_Oganesyan_Sondhi}, long-distance
properties of correlations
of $n_{\mu}(\vec{r})$ ($\mu = 0,1,2$) are well-described in this language by an operator
correspondence made up of two parts. One part, in the vicinity of the uniform wavevector
$\vec{q} = 0$, is given by lattice-gradients of $\vec{h}(\vec{R})$, and follows directly
from the construction of the microscopic heights outlined earlier, while
the other part\cite{Fradkin_Huse_Moessner_Oganesyan_Sondhi} encodes physics in the vicinity of 
wavevector $\vec{q} = {\mathbf{Q}} \equiv (\frac{2\pi}{3},-\frac{2\pi}{3})$, which
is the wavevector corresponding to columnar ordering in the honeycomb dimer
model. More explicitly, we have:
\begin{equation}
n_{0}(\vec{r})=\frac{1}{3}+\Delta_z h+c{\rm Re}(e^{i{\mathbf Q} \cdot \vec{r}} e^{{2\pi i}\overline h_0(\vec{r})}) \; ,
\end{equation}
\begin{equation}
n_1(\vec{r})=\frac{1}{3}-\Delta_y h+c{\rm Re}(e^{i{\mathbf Q} \cdot \vec{r} + \frac{2\pi i}{3}} e^{{2\pi i}\overline h_1(\vec{r})}) \; ,
\end{equation}
\begin{equation}
n_2(\vec{r})=\frac{1}{3}+\Delta_x h+c{\rm Re}(e^{i{\mathbf Q} \cdot \vec{r} + \frac{4\pi i}{3}} e^{{2\pi i}\overline h_2(\vec{r})}) \; ,
\end{equation}
where $c$ is a phenomenological constant, $\Delta_\mu h$ denotes the
lattice-gradient in the positive $\mu$ direction ($\mu=0,1,2$) along  the link of the dual triangular
lattice which cuts across the honeycomb lattice bond labeled $(\vec{r},\mu)$ (Fig.~\ref{Fig3} and Fig.~\ref{Fig4}), and ${\overline h}_{\alpha}(\vec{r})$ ($\alpha = A, B$) denotes
the average of $h(\vec{R})$ over the three triangular lattice points surrounding the honeycomb
lattice point labeled by $\alpha$ and $\vec{r}$.

In our simulations, we study the nnRVB wavefunction and the corresponding interacting dimer model on $L \times L$ lattices with periodicity of $L$
lattice units in the $\hat{e}_x$ and $\hat{e}_y$ directions. On such
periodic lattices, fully-packed dimer configurations may be labeled by three
winding numbers satisfying one constraint. The winding numbers are defined
in terms of the number of dimers $N_x$, $N_y$, and $N_z$ that
are encountered as we follow the dotted lines $X$, $Y$ and $Z$ around
the torus (Fig.~\ref{Fig4}). The definition is simply $w_{\mu} = N_{\mu} - L/3$ ($\mu = x,y,z$).
Clearly $w_x+w_y+w_z=0$ in any fully-packed configuration, since the total
number of dimers of all orientations equals $L^2$ and $N_{\mu}$ are independent
of where we cut the lattice to count these integers (both these statements
are a straightforward consequence of the fully-packed nature of each configuration).

For $L$ a multiple of $3$ (as is always the case in our numerics), $w_\mu$
are integers. For the coarse-grained height field used in our analytical effective field-theory calculations, these integers
define twisted boundary conditions, whereby $h(\vec{R})$ changes by
an amount equal to $w_\mu$ upon winding around the torus in the positive $\mu$ direction.
In a coarse-grained sense, this corresponds to a constant
gradient of strength $w_\mu/L$ in the $\mu$ direction.
Therefore, the relative statistical weight of winding sector $(w_x,w_y,w_z)$
of the honeycomb dimer model is expected to be proportional to
\begin{equation}
\exp(-\pi \kappa_d ({w_x}^2+{w_y}^2+{w_z}^2))
\label{kappa1}
\end{equation}
Thus, simply measuring the relative frequency of different winding
sectors in a Monte-Carlo simulation gives a direct handle on the phenomenological
stiffness parameter $\kappa_d$.
This is one of three ways in which we extract $\kappa_d$ in our simulations.
As we now discuss, the other two ways have to do with dimer correlations at wavevectors $\vec{q}={\mathbf Q}$ and $\vec{q}$ in the vicinity of $0$.

For wavevectors near $\vec{q}=0$, if we measure solely in the zero-winding sector (corresponding to periodic boundary conditions on $h$), we expect 
\begin{equation}
N_{0}(\vec{q}) = \langle n_0 (\vec{q}) n_0(\vec{-q}) \rangle_{0c} = \frac{1}{2\pi\kappa_d}
\left(\frac{{(q_x-q_y)}^2}{{q_x}^2+{q_y}^2+{(q_x-q_y)}^2}\right) 
\label{kappa2}
\end{equation}
\begin{equation}
N_{1}(\vec{q}) = \langle n_0 (\vec{q}) n_0(\vec{-q}) \rangle_{0c} = \frac{1}{2\pi\kappa_d}
\left(\frac{{(q_y)}^2}{{q_x}^2+{q_y}^2+{(q_x-q_y)}^2}\right) 
\label{kappa3}
\end{equation}
and
\begin{equation}
N_{2}(\vec{q}) = \langle n_0 (\vec{q}) n_0(\vec{-q}) \rangle_{0c} = \frac{1}{2\pi\kappa_d}
\left(\frac{{(q_x)}^2}{{q_x}^2+{q_y}^2+{(q_x-q_y)}^2}\right) 
\label{kappa4}
\end{equation}
in the limit $|\vec{q}| \rightarrow 0$. Here, the subscript indicates that
we take only the connected part of this correlation function and measure only in the zero-winding
sector, and we have used the
convention $\vec{q} = (q_x,q_y) \equiv q_x \hat{l}_x + q_y \hat{l}_y$ where
$\hat{l}_x$ and $\hat{l}_y$ are reciprocal lattice vectors satisfying
$\hat{l}_x \cdot \hat{e}_x = \hat{l}_y \cdot \hat{e}_y =1$ and
$\hat{l}_x \cdot \hat{e}_y = \hat{l}_y \cdot \hat{e}_x = 0$.
Each of the $N_{\mu}$ approach a limiting value of $1/(4 \pi \kappa_d)$
when $\vec{q}$ is taken to zero along two directions each: Thus, 
$N_0$ approaches this value when  $\vec{q} = (q,0)$ or $\vec{q}=(0,q)$, $N_1$ approaches this value when $\vec{q} = (q,q)$ or $\vec{q}=(0,q)$, and $N_2$ approaches this value when
$\vec{q}=(q,q)$ or $\vec{q}=(q,0)$.
This suggests that we may estimate
$\kappa_d$ rather accurately from $N(|\vec{q}|)$, the average of these six limits.

On the other hand, the connected correlation function of $n_\mu(\vec{r})$
at large spatial separation is dominated by the physics at wavevector $\vec{q} = {\mathbf Q}$. For instance, from the operator correspondence displayed earlier, we expect
\begin{equation}
\langle {n_{0}}(\vec{r})n_0(0,0)\rangle_c \sim \cos({\mathbf Q} \cdot \vec{r})\langle e^{{2\pi i}\overline h_0(\vec{r})} e^{-{2\pi i}\overline h_0(0)}\rangle \; .
\end{equation}
Evaluating the expectation value on the right-hand-side using the effective action $S$,
we find
\begin{equation}
\langle {n_{0}}(\vec{r})n_0(0,0)\rangle \sim \frac{\cos({\mathbf Q} \cdot \vec{r})}{r^{\eta_d}} \; ,
\label{kappa5}
\end{equation}
where 
\begin{eqnarray}
\eta_d &=&\frac{{\mathcal C}}{2\pi \kappa_d} \; ,
\label{kappa6}
\end{eqnarray}
with
\begin{equation}
{\mathcal C} = \int_0^{2\pi}{\frac{d\theta}{2+\sin 2\theta}} = \frac{2\pi}{\sqrt{3}}  \; . \nonumber
\end{equation}
In other words, we expect correlators at wavevector ${\mathbf Q}$ to decay
as a power-law, with power-law exponent $\eta_d = 1/(\kappa_d \sqrt{3})$; this
reflects the power-law valence-bond solid (VBS) order present in the system.
We therefore expect that a measurement of such power-law correlators at large
spatial separations provides a third independent way of extracting $\kappa_d$.

Turning to the nnRVB wavefunction, we note that the valence-bonds
in the bra define one fully-packed dimer configuration while the valence-bonds in the ket
define another fully-packed dimer configuration. The results of Ref.~\onlinecite{Damle_Dhar_Ramola} imply that these two dimer configurations have exponentially small probability
for being in two different winding sectors for large $L$. This is expected
to be true whenever the wavefunction represents a genuine spin-liquid, or,
equivalently, whenever all overlap loops are small. Since this is the case in
our wavefunction study, we restrict attention
to the sub-class of loop-model configurations in which there is no net winding of
the loops. In this restricted ensemble, which we study numerically when we sample
the nnRVB wavefunction,
the definition of winding sectors given for the dimer model goes over unchanged: We simply
keep track of the common winding numbers $w_\mu$ of the bra/ket valence-bond
configuration. This allows us to obtain an effective $\kappa_w$ for the wavefunction
directly from the analog of Eqn.~\ref{kappa1}, by simply keeping track of the histogram
of these winding numbers.

From the analysis of Ref.~\onlinecite{Damle_Dhar_Ramola}, we also expect that
the connected bond-energy correlator $D(\vec{r}) = \langle P_{\langle i j \rangle} P_{\langle k l\rangle} \rangle_c$ where $\vec{r}$ is the separtion between bonds $\langle i j \rangle$ and
$\langle k l \rangle$ and $P_{\langle i j\rangle}$ is the singlet projector on bond $\langle i j\rangle$, has the same long-distance behaviour as the connected dimer correlation of dimers living on these two bonds.
Therefore, by measuring this quantity and fitting to an oscillatory power-law
decay $\cos({\mathbf Q} \cdot \vec{r})/r^{\eta_w(g)}$, we can extract a power-law
exponent $\eta_w(g)$ for the wavefunction, and thence, an effective
stiffness parameter $\kappa_w(g) = 1/(\eta_w(g) \sqrt{3})$ exactly as in Eqn.~\ref{kappa5}
and Eqn.~\ref{kappa6} in the dimer case.

Finally, we may form the average $\bar{n}_\mu(\vec{r})$ ($\mu=0,1,2$) of the valence-bond occupation variables in the bra and the ket configuration, and consider the connected correlation functions of $\bar{n}_\mu(\vec{r})$ near wavevector $\vec{q}= 0$. From the results
of Ref.~\onlinecite{Damle_Dhar_Ramola}, we expect these to have behaviour exactly
analogous to that displayed in Eqn.~\ref{kappa2}, Eqn.~\ref{kappa3},
and Eqn.~\ref{kappa4} for the dimer model. By measuring $N(|\vec{q}|)$, the
average of the six different small $|\vec{q}|$ limits defined earlier, we expect
to obtain a third independent estimate of the effective stiffness parameter $\kappa_w(g)$ for the SU($g$) wavefunction.
These prescriptions for extracting an effective stiffness $\kappa_w(g)$ from our wavefunction simulations
are in direct correspondence with similar ideas used in Ref.~\onlinecite{Tang_Sandvik_Henley} in the square-lattice case.

\section{Algorithm}
\label{algorithm}
Our simulations of the interacting dimer model use the well-known dimer
worm algorithm of Ref.~\onlinecite{Alet_etal_PRE}. This allows
us to straightforwardly obtain high-precision results for various
dimer correlation functions even at large $L$. Our
wavefunction simulations are however much more challenging, and require some new algorithmic developments.

In order to appreciate the algorithmic difficulties involved, it is useful to
start with the following key observation, which forms the basis of
the analysis in Ref.~\onlinecite{Damle_Dhar_Ramola}: When overlap loops between bra and ket valence-bonds are on average 
rather small, as is the case in the spin-liquid phase, we may think of the corresponding
loop-gas in terms of a picture consisting of densely-packed short loops.
In the $g \rightarrow \infty$ limit where all loops are the
shortest possible, {\em i.e.} doubled-edges, we may thus caricature
the system by thinking in terms of a fully-packed dimer model where the dimers now correspond to doubled edges.

If this large-$g$ caricature of the system provides a good approximation to
long-distance properties of correlations in the nnRVB wavefunction, bra and ket
valence-bonds must necessarily be tied to each other quite strongly. This 
has important implications for the efficiency of the standard 
Sandvik-Evertz algorithm\cite{Sandvik_Evertz}. To see this, we recall
that this algorithm, when applied to wavefunction studies, consists of two steps: In the first step, one
updates the bra (ket) valence-bond
configuration using a dimer worm algorithm\cite{Alet_etal_PRE}, while keeping the ket (bra) configuration and the auxillary spin-states\cite{Sandvik_Evertz} fixed. In the second step, one 
updates the spin-states along randomly chosen overlap loops.

When one performs a dimer worm update\cite{Alet_etal_PRE} on either
the bra valence-bond configuration or the ket valence-bond configuration, the worm construction has to respect the constraints provided by the background auxillary spin-configuration, which remains static.
As a result, most worms grown by the dimer worm algorithm are extremely small,
and it is impossible to change the bra or the ket valence-bond configuration
except very slowly. This leads to serious equilibriation problems that affect the accuracy of
measurements of bond-energy correlations and winding sector probabilities in the nnRVB wavefunction
simulations for large $L$ and $g$\cite{Damle_unpublished}.

The solution\cite{Damle_unpublished} to these algorithmic difficulties
suggests itself immediately if one thinks in terms of this large-$g$ caricature for the short-loop phase: Motivated
by this caricature, one introduces
an additional update scheme, whereby a worm algorithm is used
to simultaneously move the bra and ket valence-bonds that comprise
doubled-edges in the loop representation. In other words, one considers
the sub-system made up of bra and ket valence-bonds that cover
the same link of the honeycomb lattice, {\em i.e.} the part of the lattice
which is covered by doubled-edges. This subsystem
is updated using a standard worm algorithm\cite{Alet_etal_PRE} applied to the doubled-edges. In doing so, other valence-bonds, that
form part of non-trivial overlap loops of length greater than two, are held
fixed. The spin configuration on sites visited by such nontrivial loops
is also held fixed. However, the spin labels on sites touched by doubled-edges
are updated during the construction of the worm using the following
prescription: If the worm construction starts with a $A$ ($B$) sublattice site, spin-states
of all $B$ ($A$) sublattice sites encountered in the worm construction are left unchanged,
while spin-states of all $A$ ($B$)sublattice sites encountered in the worm construction are made
consistent with the spin-state of the $B$ ($A$) sublattice site to which they are connected
by a doubled-edge in the final valence-bond configuration (after the worm has updated
that part of the lattice).

This additional update scheme greatly improves the ergodicity of our simulations.
We have tested it thoroughly in the present case and confirmed that results on small systems are identical to those obtained using the conventional Sandvik-Evertz algorithm.
At large sizes and large $g$, this additional update provides us a way
of obtaining accurate results for bond-energy correlations, and for
the relative weight of different winding number sectors. This improvement
is key to obtaining reliable results for the larger values of $g$ we
study.

\section{Numerical studies}
\label{numerics}

\begin{figure}
\includegraphics[width=\hsize]{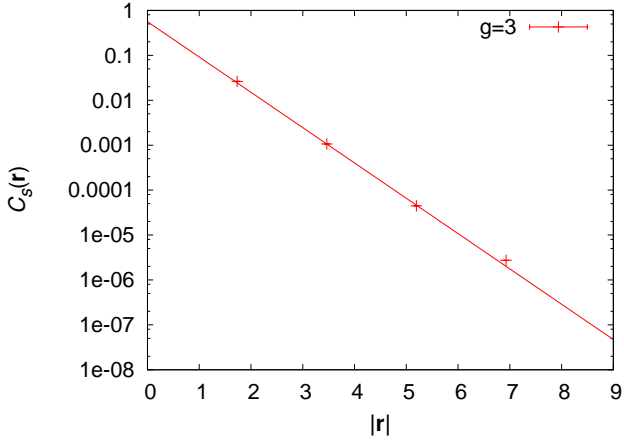}
\caption{Spin correlation function between two sites on the same sublattice, separated by $\vec{r}$.  
The fit is to an exponentially decaying function $c\exp(-r/\xi)$, with
best-fit value of correlation length $\xi=0.550(14)$.}
\label{Fig5}
\end{figure}

\begin{figure}
\includegraphics[width=\hsize]{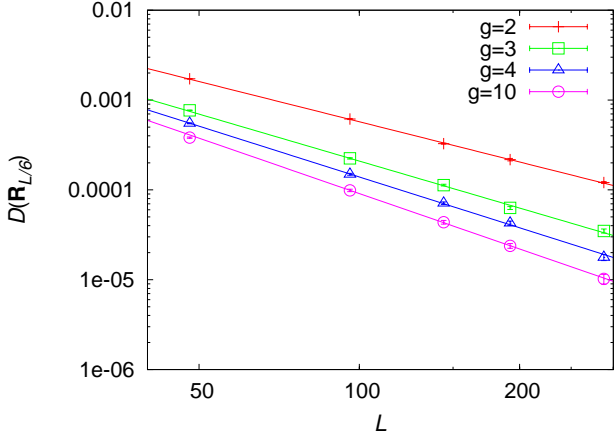}
\caption{$D(\vec{R}_{L/6})$ as a function of $L$ for
$g=2,3,4$ and $10$. $\eta_w(g)$
extracted from a fit to the form $cL^{-\eta_w(g)}$ gives $\eta_w(2) = 1.49(3)$, $\eta_w(3)= 1.74(6)$, $\eta_w(4)= 1.88(5)$, and $\eta_w(10)= 2.04(7)$. $\eta_w(g)$ is expected to equal $1/(\kappa_w(g) \sqrt{3})$, providing
us a way of estimating $\kappa_w(g)$.}
\label{Fig6}
\end{figure}

\begin{figure}
\includegraphics[width=\hsize]{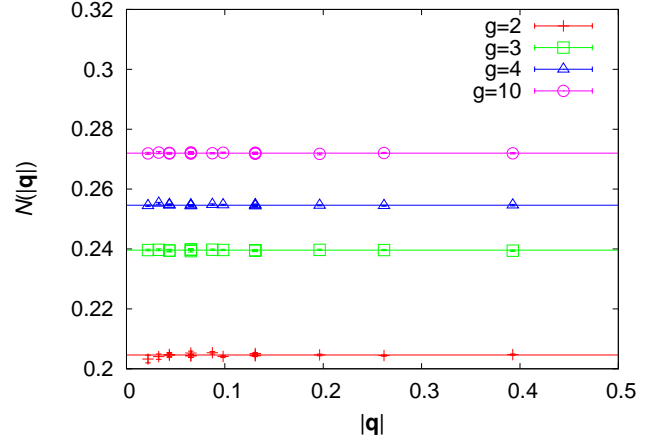}
\caption{$N(\vec{q})$ is the correlator of the average valence-bond occupation variables  (defined in Sec.~{\protect{\ref{heights}}})  in the limit of small $|\vec{q}|$, measured
in the zero-winding sector (defined in Sec.~{\protect{\ref{heights}}}) of our wavefunction simulations.
The extrapolation to $|\vec{q}| \rightarrow 0$ yields intercepts of $0.2046(2)$, $0.2396(2)$, 
 $0.2546(3)$ and $0.2722(2)$ for $g=2,3,4$ and $10$ respectively. These
intercepts are expected to equal $1/(4 \pi \kappa_w)$, and provide us an accurate estimate
of $\kappa_w(g)$.}
\label{Fig7}
\end{figure}

\begin{figure}
\includegraphics[width=\hsize]{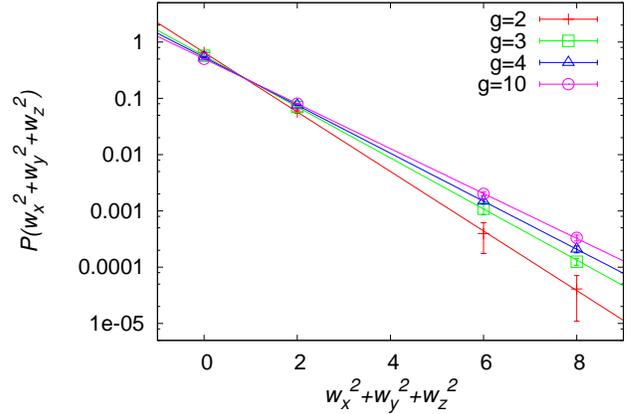}
\caption{Winding sector probabilities for the nnRVB simulation at
$g=2,3,4$ and $10$, plotted as a function of $w^2 \equiv w_x^2+w_y^2+w_z^2$. The fit is to an exponentially decaying function $a\exp{(-c(g)w^2)}$, with
best-fit values $c(2) = 1.218(5)$, $c(3)= 1.045(1)$, $c(4) = 0.982(3)$, and $c(10)= 0.919(3)$. $c(g)$
is expected to equal $\pi\kappa_w(g)$, providing a  means of estimating $\kappa_w(g)$.}
\label{Fig8}
\end{figure}

\begin{figure}
\includegraphics[width=\hsize]{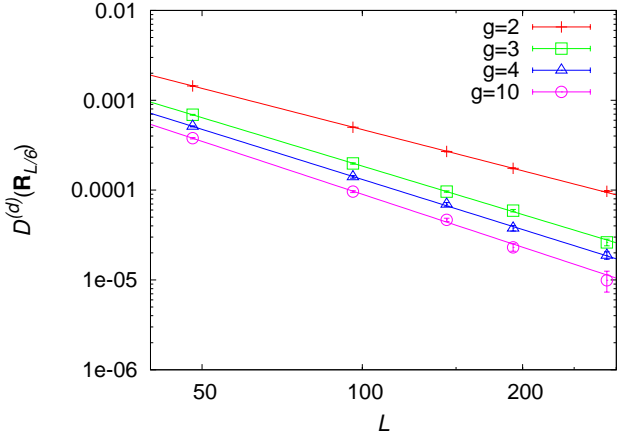}
\caption{$D^{(d)}(\vec{R}_{L/6})$ as a function of $L$ for
$g=2,3,4$ and $10$. $\eta_d(g)$
extracted from a fit to the form $cL^{-\eta_d(g)}$ gives $\eta_d(2)= 1.520(15)$, $\eta_d(3)= 1.79(2)$, $\eta_d(4)= 1.85(4)$ and $\eta_d(10)= 1.96(9)$. $\eta_d(g)$ is expected to equal $1/(\kappa_d(g) \sqrt{3})$, providing
us a way of estimating $\kappa_d(g)$.}
\label{Fig9}
\end{figure}

\begin{figure}
\includegraphics[width=\hsize]{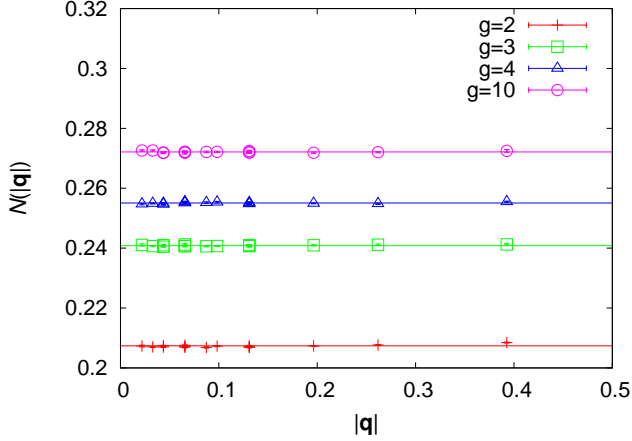}
\caption{$N(\vec{q})$ is the correlator of the average dimer correlation function (defined in Sec.~{\protect{\ref{heights}}})  in the limit of small $|\vec{q}|$, measured
in the zero-winding sector (defined in Sec.~{\protect{\ref{heights}}}) of our dimer-model simulations.
The extrapolation to $|\vec{q}| \rightarrow 0$ yields intercepts of $0.2074(1)$, $0.2408(1)$, 
 $0.2551(1)$ and $0.2721(1)$ for $g=2,3,4$ and $10$ respectively.  These
intercepts are expected to equal $1/(4 \pi \kappa_d)$, and provide us an accurate estimate
of $\kappa_d(g)$.}
\label{Fig10}
\end{figure}

\begin{figure}
\includegraphics[width=\hsize]{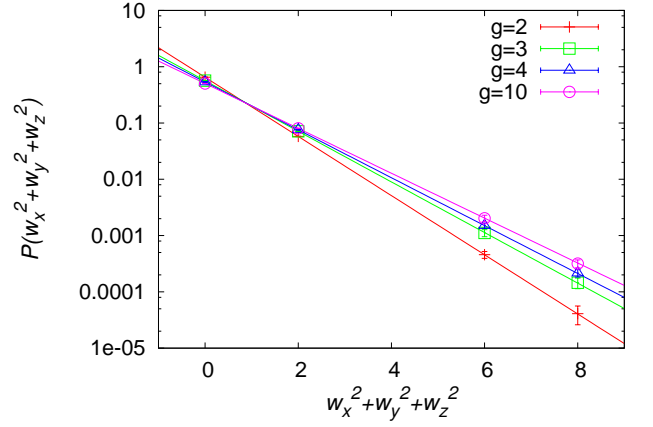}
\caption{Winding sector probabilities for the dimer-model simulation at
$g=2,3,4$ and $10$, plotted as a function of $w^2 \equiv w_x^2+w_y^2+w_z^2$. The fit is to an exponentially decaying function $a\exp{(-c(g)w^2)}$, with
best-fit values $c(2) = 1.211(4)$, $c(3)= 1.036(4)$, $c(4) = 0.979(4)$, and $c(10)= 0.918(3)$.
 $c(g)$
is expected to equal $\pi\kappa_d(g)$, providing a  means of estimating $\kappa_d(g)$.}
\label{Fig11}
\end{figure}

\begin{figure}
\includegraphics[width=\hsize]{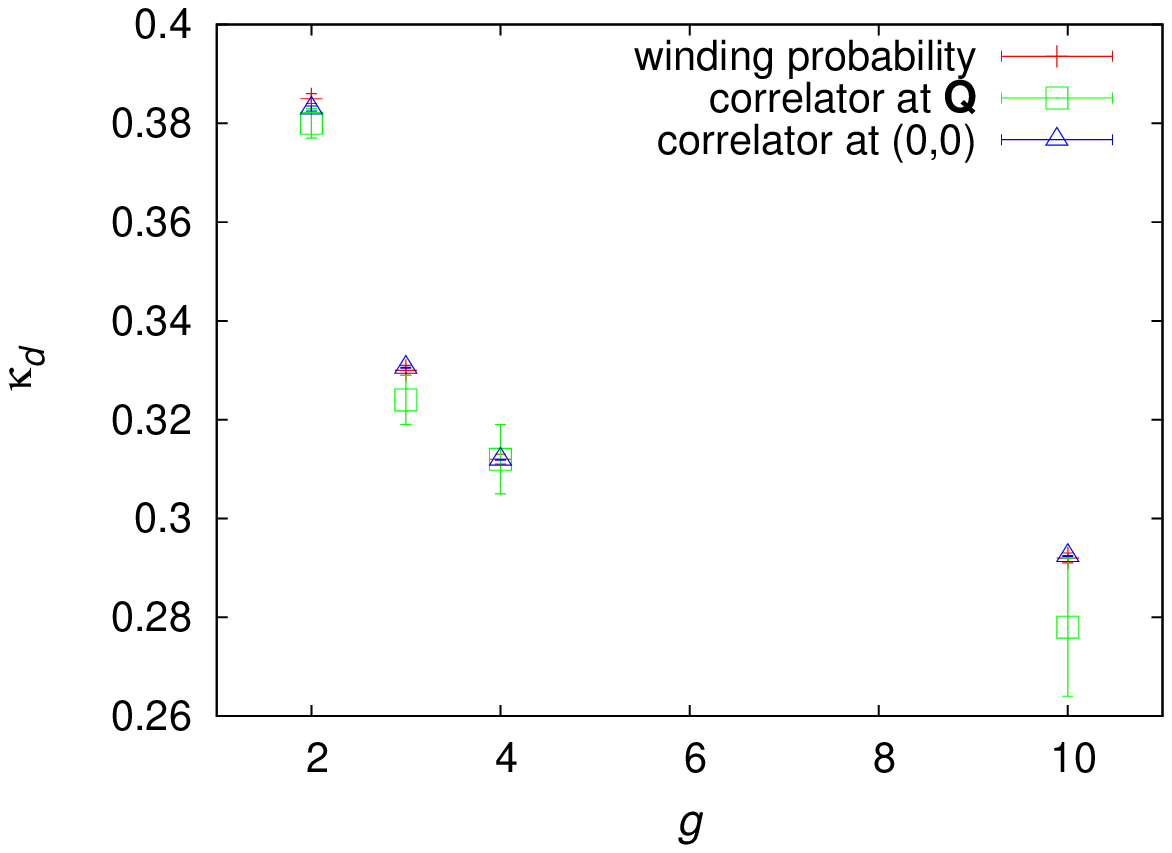}
\caption{Comparisons of values of $\kappa_d$ obtained using
different methods for $g=2,3,4$ and $10$.}
\label{Fig12}
\end{figure}

\begin{figure}
\includegraphics[width=\hsize]{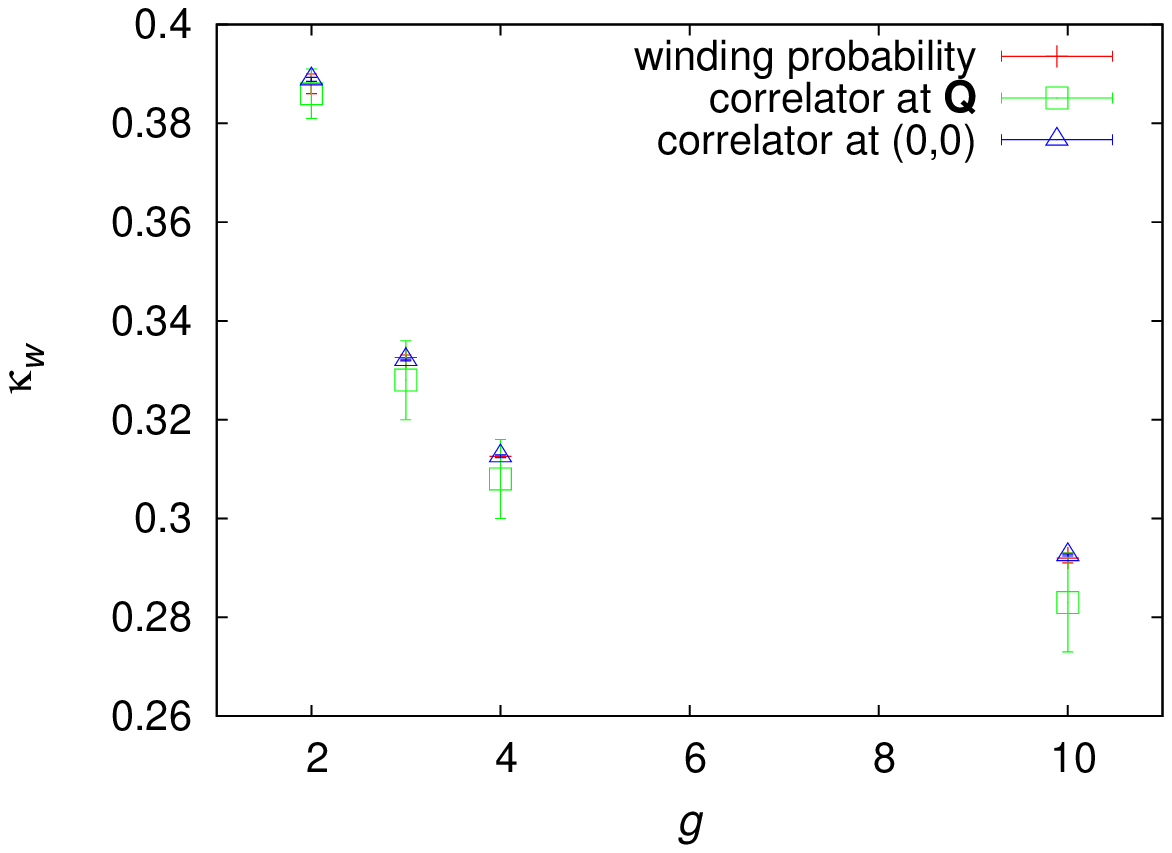}
\caption{Comparisons of values of $\kappa_w$ obtained using
different methods for $g=2,3,4$ and $10$.}
\label{Fig13}
\end{figure}

\begin{figure}
\includegraphics[width=\hsize]{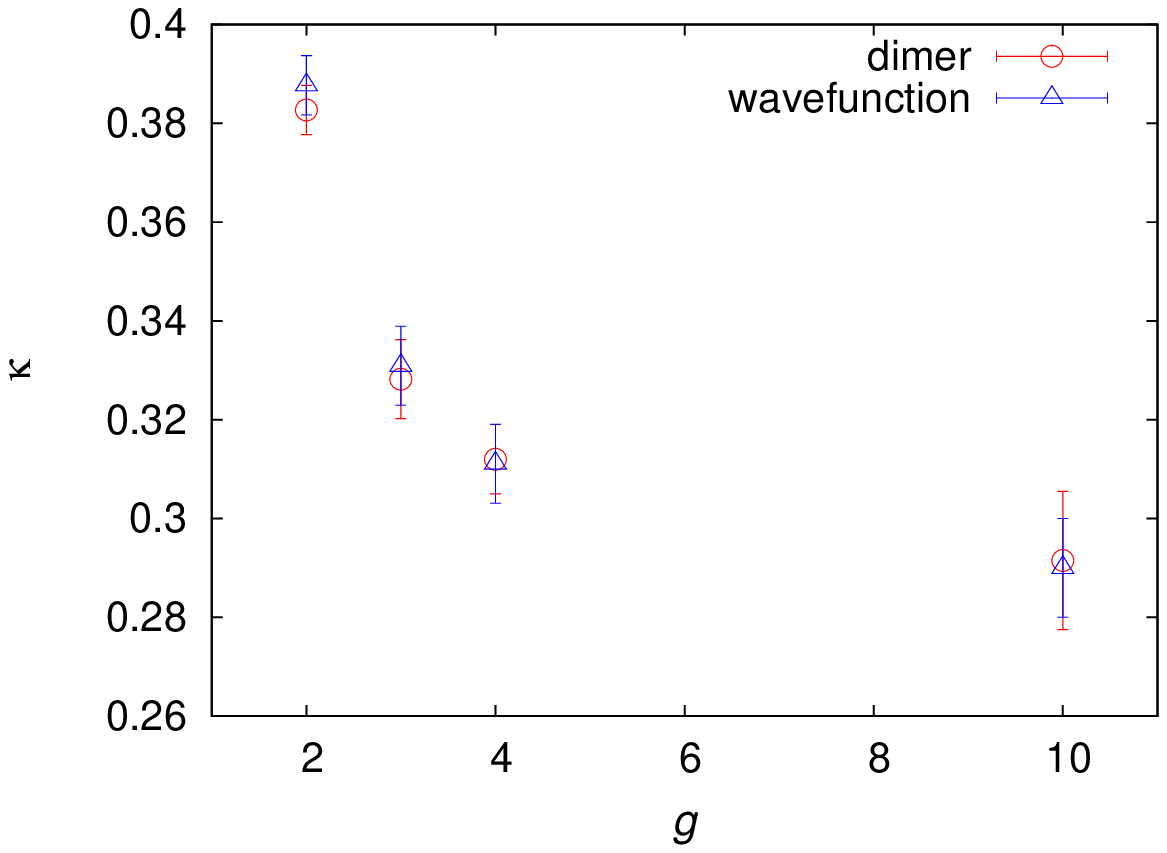}
\caption{Comparisons of the estimated values of $\kappa_w$ and $\kappa_d$,
obtained from our wavefunction and dimer model studies for $g=2,3,4,10$. 
Our consolidated estimates, considering all three ways of extracting $\kappa$
on an equal footing, are as follows: $\kappa_d (2) = 0.383(5)$, $\kappa_w (2) = 0.388(6)$, $\kappa_d (3) = 0.328(8)$, $\kappa_w (3) = 0.331(8)$, $\kappa_d (4) = 0.312(7)$, $\kappa_w (4) = 0.311(8)$, $\kappa_d (10) = 0.292(14)$, and $\kappa_w (10) = 0.290(10)$.
Error bars in these consolidated estimates reflect the spread between three different ways of estimating $\kappa_w$ and $\kappa_d$, as well as statistical errors in the individual estimates.}
\label{Fig14}
\end{figure}

In our wavefunction studies,
we study $L\times L$ systems of $2L^2$ spins(Fig~\ref{Fig3}),
where $L$ a multiple of $12$ ranging from $L=48$ to $L=288$. 
To characterize the N\'eel order in the ground state, it is conventional to construct
the corresponding order parameter 
\begin{equation}
\vec{M}_s = \frac{1}{L^2}\sum_{\vec{r}} \vec{m}(\vec{r})
\end{equation}
where $\vec{m}$ is the local N\'eel order parameter field defined as
\begin{equation}
\vec{m}(\vec{r}) = \vec{S}_{\vec{r} A} - \vec{S}_{\vec{r}B}
\end{equation}
where $\vec{r}A$ ($\vec{r}B$) refers to the $A$ ($B$) sublattice
site ``belonging" to Bravais lattice site $\vec{r}$ (Fig.~\ref{Fig3}).
We have studied the $L$ dependence of $\langle \vec{M}_s^2\rangle$ in
the SU($g$) wavefunction and confirmed that there is no long-range
N\'eel order for any $g$. The short-ranged nature of spin correlations
are particularly obvious in our results for the spin correlation
function:
\begin{equation}
C_s(\vec{r})=\frac{{\cal A}_g}{L^2}\sum_{\vec{r}'}
\langle\vec{S}_A(\vec{r}'+\vec{r}) \cdot \vec{S}_A(\vec{r}')\rangle
\end{equation}
Here, $\vec{S}$ are spin-half operators in the $g=2$
case; more generally, they are spin $S=(g-1)/2$ operators in the SU($g$) case. The normalization ${\cal A}_g = 3/(g^2-1)$ is inserted
to yield a Monte-Carlo estimator\cite{Sutherland,Beach_Sandvik,Beach_Alet_Mambrini_Capponi,Lou_Sandvik_Kawashima} that takes
on a $g$-independent value for a given configuration of loops.

To characterize the tendency towards
power-law VBS order at the columnar wavevector ${\bf Q}$, we define the
columnar VBS order parameter $\Psi = \sum_{\vec{r}} V_{\vec{r}}$, where $V_{\vec{r}}$ is the local VBS order parameter field defined as:
\begin{equation}
V_{\vec{r}} = (P_{\vec{r}0} + e^{2 \pi i /3} P_{\vec{r} 1} + e^{4 \pi i/3} P_{\vec{r} 2})e^{i \vec{Q} \cdot \vec{r}} \; .
\end{equation}
Here $P_{\vec{r} \mu}$ ($\mu = 0, 1,2$) denotes the SU($g$) singlet projector
on the bond labeled by $\mu$ and $\vec{r}$ (Fig.~\ref{Fig3}),
and ${\mathbf Q} \equiv (2\pi/3,-2\pi/3)$ (Fig.~\ref{Fig3}).
In our numerical work, we find it convenient to
compute correlations of $V_{\vec{r}}$
\begin{equation}
D(\vec{R}_{\alpha L}) = \frac{{\cal B}_g}{L^2}\sum_{\vec{r}} \langle V^{\dagger}_{\vec{r}+\vec{R}} V_{\vec{r}} + V^{\dagger}_{\vec{r}} V_{\vec{r}+\vec{R}} \rangle 
\end{equation}
at separations $\vec{R}_{\alpha L} = (\alpha L, \alpha L)$ with $\alpha = \frac{1}{3},
\frac{1}{4}, \frac{1}{6}$ . 
Here, ${\cal B}_g = 9g^4/(16(g^2-1)^2)$ is a normalization
introduced to ensure that the {\em leading contribution} to the estimator\cite{Sutherland,Beach_Sandvik,Beach_Alet_Mambrini_Capponi,Lou_Sandvik_Kawashima} for this correlation function is independent
of the value of $g$ (this leading contribution comes from loop configurations in which
$\vec{r}$ and its neighbour both lie in one loop, and $\vec{r}+\vec{R}$ and its neighbour
both lie in another loop).
If the correlator of $V_{\vec{r}}$ decays as $1/r^{\eta_w(g)}$, we expect
this to be reflected in a $1/L^{\eta_w(g)}$ decay of $D(\vec{R}_{\alpha L})$
for fixed $\alpha$.

Fig.~\ref{Fig5} displays the exponential decay of the spin-correlation function,
with correlation length of order half the spacing between two consecutive
$A$ sublattice sites. In sharp contrast to this behaviour,
we find that the correlator of $V_{\vec{r}}$ decays as a slow power-law
$\sim r^{-\eta_w(g)}$ 
where $\eta_w(g)$ is an increasing function of $g$. This is clear from
Fig.~\ref{Fig6}, which shows the $L$ dependence of $D(\vec{R}_{\alpha L})$
for $\alpha = 1/6$.  As noted earlier,  we may extract an
effective stiffness $\kappa_w(g)$ from this power-law exponent $\eta_w(g)$
via the relation $\kappa_w(g) = 1/(\eta_w(g) \sqrt{3})$.
In Fig.~\ref{Fig7}, we show the $|\vec{q}|$ dependence of $N(|\vec{q}|)$, the average
valence-bond correlator defined earlier in Sec.~\ref{heights} (averaged over six
quantities, all of which are expected to tend to the limit $1/(4 \pi \kappa_w)$ in
the small $|\vec{q}|$ limit) and measured in the zero-winding sector.
From this limiting behaviour of $N(|\vec{q}|)$, we are able to extract our most
accurate estimate of $\kappa_w(g)$.
Finally, we display winding sector probabilities (in the restricted ensemble in
which both bra and ket valence bonds have the same winding numbers) from
our wavefunction simulations in Fig.~\ref{Fig8}. By fitting these to
an exponentially decaying function of $w^2 \equiv w_x^2+w_y^2+w_z^2$, we
obtain a third independent estimate of $\kappa_w(g)$.

In order to test the correspondence between bond-energy correlations
in the SU($g$) wavefunction and dimer correlations in the interacting
dimer model, we characterize VBS order in the interacting dimer model
in a completely analogous way, in terms of the columnar VBS order parameter $\Psi_d = \sum_{\vec{r}} V^{(d)}_{\vec{r}}$, where $V_{\vec{r}}^{(d)}$ is the local VBS order parameter field defined as:
\begin{equation}
V_{\vec{r}}^{(d)} = (n_{0}(\vec{r}) + e^{2 \pi i /3} n_{1}(\vec{r}) + e^{4 \pi i/3} n_{2}(\vec{r}))e^{i \vec{Q} \cdot \vec{r}} \; .
\end{equation}
Here $n_{\mu}(\vec{r})$ ($\mu = 0, 1,2$) denotes the dimer occupation number
on the bond labeled by
$\mu$ and $\vec{r}$ (Fig.~\ref{Fig3}),
and ${\mathbf Q} \equiv (2\pi/3,-2\pi/3)$ (Fig.~\ref{Fig3}).

To probe the VBS order in the system, we compute correlations of $V_{\vec{r}}^{(d)}$
\begin{equation}
D^{(d)}(\vec{R}_{\alpha L}) = \frac{1}{L^2}\sum_{\vec{r}} \langle (V^{(d)}_{\vec{r}+\vec{R}} )^{\dagger}V^{(d)}_{\vec{r}} + (V^{(d)}_{\vec{r}} )^{\dagger}V_{\vec{r}+\vec{R}}^{(d)} \rangle 
\end{equation}
at separations $\vec{R}_{\alpha L} = (\alpha L, \alpha L)$ with $\alpha = \frac{1}{3}, \frac{1}{4}, \frac{1}{6}$. As in the wavefunction case, we expect  $D^d$ to decay
as $1/L^{\eta_d(g)}$  when correlations of $V_{\vec{r}}^{d}$ decay as $1/r^{\eta_d(g)}$.
In Fig.~\ref{Fig9}, we see that this is indeed the case. From power-law fits
to this behaviour, we obtain $\eta_d(g)$, and thence, an estimate for
$\kappa_d(g) = 1/(\eta_d(g) \sqrt{3})$.
In Fig.~\ref{Fig11}, we show our data in the zero-winding sector for the average $N(|\vec{q}|)$ of
the six dimer correlators that are all expected to approach $1/(4 \pi \kappa_d)$
in the limit of small $|\vec{q}|$ (as discussed in Sec.~\ref{heights}). From the
limiting behaviour of this function, we obtain a very accurate estimate for
$\kappa_d(g)$. Finally, we display the relative probabilities for different
winding sectors in Fig.~\ref{Fig11}. Fitting this to an exponentially decaying
function of $w^2 \equiv w_x^2+w_y^2+w_z^2$ gives us a third independent estimate of $\kappa_d(g)$.

Fig.~\ref{Fig12} and Fig.~\ref{Fig13} show the values obtained for $\kappa_d(g)$
and $\kappa_w(g)$ in these three ways. As is clear from these figures, all three
ways of extracting a stiffness are in rather good mutual agreement both for the wavefunction,
and for the interacting dimer model. This is strong evidence for the correctness
of the coarse-grained height-description in both cases. Finally, in Fig.~\ref{Fig14},
we display the average of the three estimates for $\kappa_w(g)$ compared
with the corresponding average of estimates for $\kappa_d(g)$. As is clear
from this figure, the interacting dimer models studied at each $g$
provide a remarkably good quantitative account of the long-distance properties
of the SU($g$) wavefunction. This is our main result.

\section{Discussion}
\label{discussion}
Does this correspondence with an interacting dimer model continue
to provide useful insights when the
RVB wavefunction has longer-range bipartite valence bonds while preserving
the Marshal sign-structure\cite{Akbar1} on the honeycomb
lattice? The answer is clearly yes,
although the form of the interactions gets correspondingly more complicated.
What about more complicated wavefunctions which also have a non-trivial sign-structure (in the $S^z$ basis), as is expected to be the case in the ground-state\cite{Akbar2,honeycomb_DQCP} of the honeycomb-lattice Heisenberg
model with frustrating further neighbour couplings?
The answer is much less clear since the non-trivial sign structure would
necessarily lead to a description with Boltzmann weights carrying non-trivial phase-factors when expressed in the valence-bond basis. In this context, it is perhaps useful to note that certain dimer models with such
general Boltzmann weights have recently been solved on the square
lattice\cite{Aiyyer}. In another recent strand of work\cite{Wildeboer_Seidel,Yang_Yao}, it has also been shown that 
a class of nnRVB wavefunctions on frustrated planar lattices can be rewritten in the $S^z$ basis as a sign-free partition sum with ``Boltzmann weights'' expressed in terms of Pfaffians, thereby allowing
efficient Monte-Carlo calculation of physical observables. It would be interesting
to ask if these Boltzmann weights have a controlled expansion in terms of some
classical spin model with tractable interactions.
What about the three-dimensional case? In the isotropic
case\cite{Albuquerque_Alet_Moessner}, the framework used here is of very limited utility since overlap loops are long
and the nnRVB wavefunction has N\'eel order. However, the present
framework should be able to provide some insights into the nature of
the spin-liquid phases explored recently in anisotropic three-dimensional RVB wavefunctions\cite{Xu_Beach_unpublished}.

\section{Acknowledgements}
One of us (KD) would like to thank F.~Alet for useful discussions about the results
of Ref.~\onlinecite{Albuquerque_Alet}.
This project was initiated under the umbrella of the Gulmohar Center (IIT Bombay), while the final stages of this project
were supported by the Visiting Student Research Program of the TIFR. The authors are grateful to both
institutions for making this collaboration possible. This research was
supported by the Indo-French Centre for the Promotion of Advanced Research (IFCPAR/CEFIPRA) under Project 4504-1, and used computational resources funded by
DST-SR/S2/RJN-25/2006 in addition to departmental computational
resources of the Dept. of Theoretical Physics of the TIFR.

\end{document}